\documentclass[a4paper,11pt]{article}
\usepackage{graphics}

\makeindex

\begin{document}

\title{An analytical analysis of  vesicle tumbling under a shear flow}

\author{\lineskip 1em
F. Rioual, T. Biben and C. Misbah,\\
{\sl Groupe de Recherche sur les Ph\'enom\`enes Hors Equilibres, 
L.S.P.}\\
{\sl Universit\'e Joseph Fourier, CNRS (UMR 5588),}\\
{\sl B.P. 87, F-38402 Saint Martin d'H\`eres cedex, France }
}
\date{}
\maketitle

\begin{abstract}
Vesicles under a shear flow exhibit a tank-treading motion of their
membrane, while their long axis points with an angle $<\frac{\pi}{4}$ with respect 
to the shear stress if the viscosity contrast between the interior and the
exterior is not large enough. Above a certain viscosity contrast, the
vesicle undergoes a tumbling bifurcation, a bifurcation which is known 
for red blood cells. We have recently presented the full numerical analysis of this transition.
In this paper, we introduce an 
analytical model that has the advantage of being both simple enough and capturing 
the essential features found numerically. The model is based on general 
considerations and does not resort to the explicit computation of the 
full hydrodynamic field inside and outside the vesicle.
\end{abstract}

\hspace{.3in}{\sl PACS numbers: 87.16.Dg, 47.60.+i, 87.17.Jj} %

\section{Introduction}

Vesicles are closed membranes, which are suspended in an aqueous 
solution.
They represent an attractive biomimetic system, which has revealed 
several interesting static and dynamical features that bear a strong 
resemblance with some behavior of real cells. Among these features we can cite
equilibrium shapes \cite{Lipowsky} revealing forms similar
to red blood cells, and tumbling known for these cells \cite{Goldsmith}. It 
is known that red cells, like vesicles \cite{Kraus}, orient
themselves at a given angle with respect to the shear flow at high
hematocrit (high enough concentration of red cells), while at low 
hematocrit (where cells behave as being individual) both in vitro and in vivo
observations reveal a tumbling motion, where the long axis of the red 
cell rotates in a periodic fashion. It has been recognized for a long time
that the viscosity ratio between the internal fluid and the ambient 
one
is a decisive factor (the more viscous is the internal fluid in 
comparison to the external one, the easiest is the tumbling). Another relevant 
ingredient is
the swelling ratio: a flatten out cell would tumble more easily than a
swollen one. Several attempts in understanding the tumbling transition 
have been made in the litterature, the most prominent one is the work of 
Keller and Skalak \cite{keller}. This work uses the solution of the 
hydrodynamical equations in the Stokes regime (inertial effects are negligibly 
small for biological blood transport) around an ellipsoid which involve quite 
complex expressions. But still several assumptions had to be made in order to 
solve the problem. Recently, a full numerical analysis has been presented 
\cite{beaucourt03} and provided the boundaries in the parameter space 
(basically the viscosity contrast, and the swelling ratio) separating
the regions of tumbling and those of tank-treading.
That work focused on vesicles
that correspond to a simplified model of red blood cells, and especially
ignored the elastic properties of the membrane, a fact which though 
turns out to lead to some interesting qualitative changes, will not be accounted 
for here either.

Due to the interplay of several effects in the tumbling transition, it 
is highly desirable to have at our disposal an analytical theory, which, on 
the one hand, should reproduce the basic essential features of the tumbling
transition, and on the other, should be simple enough in order to shed light on
the various competing phenomena leading to tumbling. It is the
main aim of the present paper to deal with this question.

The present theory bypasses the tedious computation of the velocity 
field around the ellipsoid (Note that in the general case, no analytical solution of the Stokes flow
 is known), and is based on the assumption that the forces
acting on each piece of the vesicle membrane are proportional to the 
actual relative velocity at the membrane with respect to the applied flow.
It follows from our study that simple enough
notions account remarkably well for many features and render each 
effect transparent. In addition, this work offers a promising basis for more
elaborate models, including, for example, the effect of membrane 
stretching or shear elasticity.

The scheme of this paper is as follows. In section \ref{ingredient}, we
present the basic ingredients of the model. Section \ref{moment} is 
devoted to the derivation of the dynamical equation that governs the motion of 
the vesicle. This part is based on a torque balance. Section \ref{vtank} 
presents a complementary ingredient that serves to put the evolution equation in 
a closed form. This is based on an energy balance between
the inner fluid of the vesicle and the work provided by the ambient one. The main outcomes 
of the analytical theory together with their confrontation with the full numerical analysis 
are presented in section \ref{outcome}.

\section{Basic ingredients of the model}
\label{ingredient}

\begin{figure*}[tbp]
\begin{center}
\scalebox{.6}{\rotatebox{0}{\includegraphics*{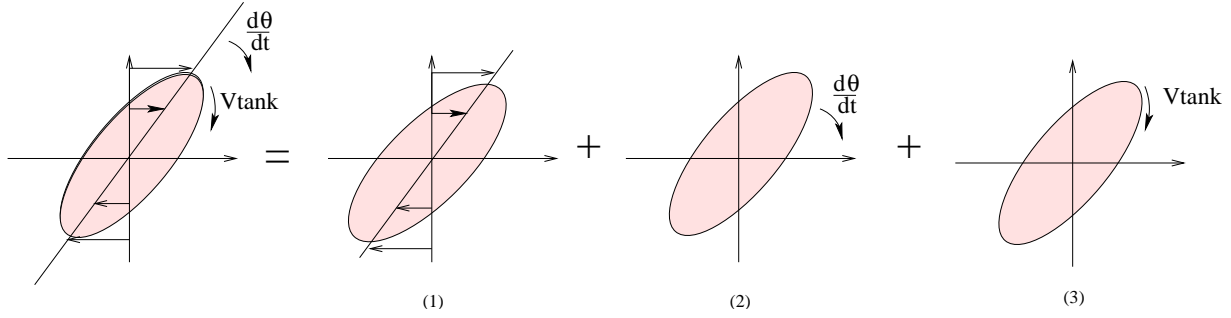}}}
\end{center}
\caption{Decomposition of the velocity field around the vesicle 
subjected to a tumbling motion and a tank-treading motion of its 
membrane in a simple shear flow.}
\label{figflow}
\end{figure*}

\begin{itemize}

\item As in \cite{keller}, we will assume the shape of the vesicle to be
an undeformable ellipse, with the long and the short axes denoted by $a$ 
and $b$, respectively. 
It will be recognized that the theory can be used 
for arbitrary prescribed shapes. For definitness11 we shall, however, specialize our discussion to  an 
elliptical shape. The enclosed area is denoted by $S$, and the perimeter 
by $P$. The fluid embedded into the vesicle has a viscosity
$\mu_{in}$ and the ambient one $\mu_{out}$. $\tau$ is the swelling ratio of
the vesicle defined in $2D$ as $\tau =4 \pi S/P^2$. For a circle 
$\tau=1$ and it is smaller than one otherwise.

\item The vesicle is subjected to a linear shear flow $(v_{x}=\gamma
y, v_{y}=0)$ where $\gamma$ is the shear rate. Since the vesicles of
interest have a fluid membrane, each material point on the membrane will 
be
transported by the flow, so that the membrane moves in a tank-tread 
fashion.

Let us make a remark, which will prove to be useful later in this paper. 
A
simple shear flow characterized by the (2$\times$2) shear rate matrix

\[\left
(\begin{array}{cc}
0 & \gamma \\
0 & 0
\end{array}
\right) \]

can be decomposed into two parts: a symmetric one given by
\[\left
(\begin{array}{cc}
0 & \gamma /2 \\
\gamma /2 & 0
\end{array}
\right) \]

and an antisymmetric one given by

\[\left
(\begin{array}{cc}
0 & \gamma /2 \\
-\gamma /2 & 0
\end{array}
\right) \]

As shown on Fig. \ref{dshear}, the antisymmetric part provides a 
rigid-like clockwise rotation of the vesicle (R), while
the symmetric part corresponds to an elongational (or strain) flow, which 
tends to orient the vesicle along $\frac{\pi}{4}$ (E).

\begin{figure*}[tbp]
\begin{center}
\scalebox{.5}{\rotatebox{0}{\includegraphics*{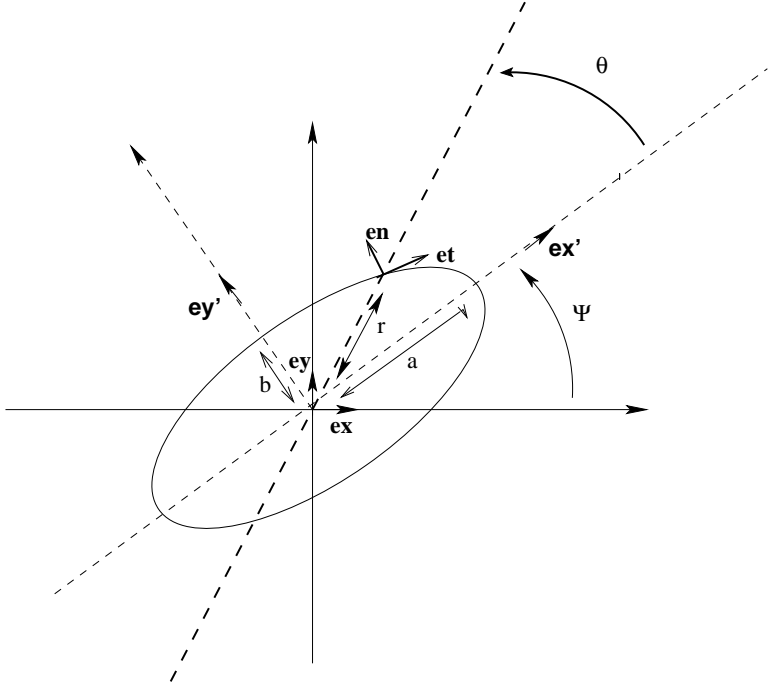}}}
\end{center}
\caption{The different frames involved in the model}
\label{axis}
\end{figure*}

\begin{figure*}[tbp]
\begin{center}
\scalebox{.5}{\rotatebox{0}{\includegraphics*{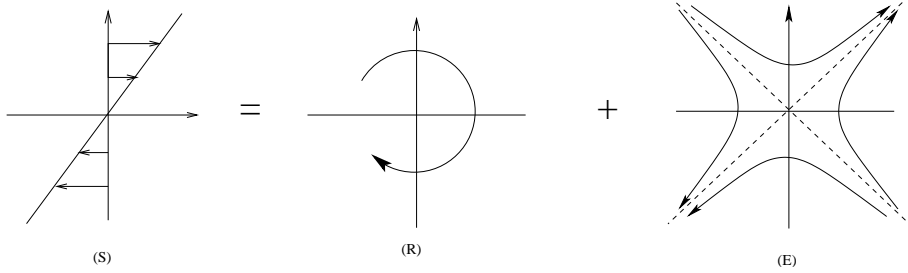}}}
\end{center}
\caption{Decomposition of the shear flow (S) in a rotational part (R) and an 
elongational part (E).}
\label{dshear}
\end{figure*}

Our calculation is based on the following two properties of the Stokes
equations:

\item Due to the linearity of the Stokes equations, the superposition
principle for given boundary conditions applies: the velocity field 
around a
vesicle subjected to a tank-treading and a tumbling motion in a simple 
shear
flow is the sum of the velocity fields obtained for the three following 
configurations (see Fig. \ref{figflow}):

\begin{itemize}
\item {A simple shear flow acting on a rigid body fixed in the flow at a 
constant orientation angle $\psi$
with a fluid velocity equal to zero on the contour of the vesicle.}

\item {The flow created by a rigid elliptic body rotating at a rotation
velocity $\frac{d\psi}{dt}$ in a quiescent fluid.}

\item {The flow created by an elliptic body subjected to a tank-treading
motion of its contour and fixed at a constant orientation $\psi$ in a 
quiescent
fluid.}
\end{itemize}

\item The second ingredient, which follows from the previous one, is an 
extension of a general result valid in Stokes flows for a solid which is 
in relative motion at a velocity $V$ with
respect to the surrounding fluid. The drag force on the solid scales as 
$F_{drag}\hspace{.1in}=\mu\hspace{.1in}\lambda\hspace{.1in}V$,
 where $\lambda$ is (a drag coefficient) function of the geometry of the body.
\footnote{More precisely, the Stokes force exerted on a solid of typical
length $L$ in a
translational motion at speed $\mathbf{U}$, in a quiescent newtonian fluid of 
viscosity $\mu$,
scales indeed as
\[
F\approx \mu UL
\]
More formally, we can write a linear relation between the force and the
velocity:
\begin{equation}
F_{i}=-\mu A_{ij}U_{j}
\end{equation}
$A_{ij}$ is a tensor which is symmetrical for a newtonian fluid, and in 
a specific frame linked
to the solid, one can write :
\begin{equation}
F_{i}=-\mu \lambda_{i}U_{i}
\end{equation}
}
There is a linear relation between the force and the relative velocity
 of the body with respect to the applied flow. We view the elliptic
contour as being represented
by adjacent segments. The key hypothesis of the following analysis is to
apply this property,
ie. the linearity between forces and relative velocities, on each 
segment of the membrane.
Let us make some important comments about the meaning of this assumption.
The external force applied on an elementary segment of the membrane is 
provided, on the one hand, by
 the flow imposed externally, and, on the other hand, by the back-flow 
due to the presence of the  vesicle.
 This retroaction of the vesicle on the applied flow is a
 complex piece of the study and an exact determination
of its effect requires sophisticated numerical treatments such
as the Boundary Integral method (\cite{cantat99,pozrikidis}).
In our model, the basic assumption stated above takes into account this 
complex interaction in an effective manner:
the effect of the back-flow is included in the  coefficient $\lambda$, 
which links the effective force to the relative velocity of the segment 
with respect to the applied flow.
In the framework of our model, this  coefficient is chosen to be 
independant on the particular
elementary segment considered. This is reminiscent of a ''mean-field''
like approximation. 
This  coefficient is also a priori non isotropic, {\it i.e.} takes different
values depending on wether we consider the
normal direction or the transverse direction of the elementary segment
considered.
These two values will be denoted as $\lambda_t$ and $\lambda_n$ and 
their determination will be discussed in section \ref{outcome}. To some 
extent this model is akin to the Rouse model for polymer rods \cite{Doi}, 
where hydrodynamical interactions between adjacent pieces are ignored.
\end{itemize}

\section{Mechanical equilibrium for the vesicle in the shear flow}

\label{moment}

As stated above, the main idea is to use the linear generalized Stokes 
law at the local level
of each segment of the contour, and to compute the torque associated with 
the
force. Since we shall decompose the velocity field into an applied 
shear, a
tank-treading motion, and a tumbling one, we shall have to deal with 
three
types of forces separately. Once each torque is evaluated, we sum up the
three contributions, and set the resultant to be zero, owing to absence
of
inertia. Once the expressions of the forces are specified, the remaining
pieces of the work are purely algebraic with some specific integrals
involving the geometry of the vesicle.

Since the Stokes law relating the force to the relative velocity is 
local, we
find it convenient to first, write it in the frame linked to the vesicle, 
and then, to express the torque elements in the laboratory frame for ease of
computations. We refer to Fig. \ref{axis} for the different frames used 
here.
The laboratory frame has the basis denoted as $\mathbf{(e_x,e_y)}$. The 
rotating frame,
which is linked to the principal axes of the ellipse, is specified
by its basis denoted  $\mathbf{(e_x^{\prime},
e_y^{\prime})}$. The local frame associated to an elementary segment on 
the
elliptic contour is specified by $\mathbf{(e_t,e_n)}$.

In the local frame $\mathbf{(e_t,e_n)}$, the components of the local 
force (or drag)
applied on the segment per unit length in the transverse direction can 
be
expressed as functions of the relative velocities $(V_{t},V_{n})$ exerted upon
a membrane element:

\begin{equation}
\begin{array}{l}
dF_{t}= -\mu_{out} \lambda_{t} V_{t} dl\\
dF_{n}= -\mu_{out} \lambda_{n} V_{n} dl
\end{array}
\label{lforces}
\end{equation}
where $\lambda_t$ and $\lambda_n$ are phenomenological parameters of the
model associated respectively to the transverse and the normal motions 
to
the segment $dl$. $\lambda_t$ and $\lambda_n$ have positive values with 
the choice of Eqs. \ref{lforces}.
They have a dimension of the inverse of a length.
The crux of the analysis is to decompose the local velocity in three 
pieces as stated above, and
evaluate various torques.

\begin{itemize}
\item {\emph{(1): Torque of the force acting on the body in a simple 
shear
flow}}

In the laboratory frame $\mathbf{(e_x,e_y)}$, the velocity field of a
simple shear flow takes the form:

\[
\begin{array}{c}
v_{x}=\gamma\: y \\
v_{y}=0
\end{array}
\]
$\gamma$ is the  shear rate, which fixes the time-scale of the flow.
Written in the local coordinate system $\mathbf{(e_t,e_n)}$, the relative 
velocity reads:

\[
\begin{array}{l}
V_{shear_t} = -\gamma\: y \mathbf{e_x.e_t}\\
V_{shear_n} = -\gamma\: y \mathbf{e_x.e_n}
\end{array}
\]

Using (\ref{lforces}), we determine the associated forces denoted as $
dF_{shear_t}$ and $dF_{shear_n}$, from which the torque is computed as:



\begin{equation}
\mathbf{M_{shear}}=\oint_{C}\mathbf{r}\times \mathbf{dF_{shear}}
\label{Mshear}
\end{equation}
Using the coordinates linked with the natural axes of the ellipse (for a 
convenient calculation),
we easily find:

\begin{equation}
M_{shear}=\mu_{out}\gamma \left[\frac{L_{s1}^{2}+L_{s2}^{2}}{2}+
\frac{L_{s1}^{2}-L_{s2}^{2}}{2} \cos (2\psi) \right]
\label{shear}
\end{equation}
with the convention that a positive torque corresponds to a clockwise
rotation (see Fig. \ref{axis}). $L_{s2}$ and $L_{s1}$ are elliptic integrals:

\[
\begin{array}{c}
L_{s2}^{2}=\left[\lambda_{t}f(a,b)-\lambda_{n}f(b,a)\right] + 
(\frac{b}{a})^{2}\left[(\lambda_{t}g(a,b)+\lambda_{n}f(b,a)\right] \\
L_{s1}^{2}=\left[\lambda_{t}f(a,b)-\lambda_{n}f(b,a)\right] + 
(\frac{a}{b})^{2}\left[(\lambda_{t}g(b,a)+\lambda_{n}f(b,a)\right]
\end{array}
\]

with

\[
\begin{array}{c}
f(a,b)=\oint_{C}\frac{x^{\prime }{}^{2}y^{\prime }{}^{2}}{(\frac{a}{b}
)^{2} y^{\prime }{}^{2}+(\frac{b}{a})^{2} x^{\prime }{}^{2}}dl \\
g(a,b)=\oint_{C}\frac{x^{\prime }{}^{4}}{(\frac{a}{b})^{2} y^{\prime
}{}^{2}+(\frac{b}{a})^{2} x^{\prime }{}^{2}}dl
\end{array}
\]

According to the linear decomposition of Fig. \ref{dshear}, we can 
identify from (\ref{shear})
the torque associated
with the rotational part of the flow:

\begin{equation}
M_{rot}= \mu_{out}\gamma\frac{L_{s1}^2+L_{s2}^2}{2}
\label{Mrot}
\end{equation}

and the torque associated with the elongational part of the flow:

\begin{equation}
M_{elong}= \mu_{out}\gamma\frac{L_{s1}^2-L_{s2}^2}{2}\cos (2\psi)
\label{Melong}
\end{equation}

\item {\emph{(2) Torque of the force acting on a rigid ellipse with a
rotation speed $\frac{d\psi}{dt}$}}\newline

The tumbling velocity at a position $r$ of the membrane is given by:

\begin{equation}
\mathbf{V_{tumble}}=\mathbf{w}\times\mathbf{r}
\end{equation}

where $\mathbf{w}=\frac{d\psi}{dt}\mathbf{e_z}$ is the instantaneous angular
velocity of the vesicle $\frac{d\psi}{dt}$.




and its components in the local frame are

\[
\begin{array}{c}
V_{tumble_{t}}=(x^{\prime}\frac{d\psi}{dt}\mathbf{e_{y^{\prime}}}
-y^{\prime}\frac{d\psi}{dt}\mathbf{e_{x^{\prime}}}) \\
V_{tumble_{n}}=(+x^{\prime}\frac{d\psi}{dt}\mathbf{e_{y^{\prime}}}
-y^{\prime }\frac{d\psi}{dt}\mathbf{e_{x^{\prime}}})
\end{array}
\]
%


We use (\ref{lforces}) to determine the force and then we compute
the torque as:

\[
\mathbf{M_{tumble}}=\oint_{C}\mathbf{r}\times \mathbf{dF_{tumble}}.
\]
This yields, after elementary integration, to:

\begin{equation}
M_{tumble}=\mu _{out}\frac{d\psi}{dt}(L_{s1}^{2}+L_{s2}^{2})
\label{Mtumble}
\end{equation}

\item {\emph{(3) Torque of the force acting on the ellipsoid related to 
the
tank-treading motion}}

The tank-treading velocity is tangential to the membrane:

\begin{equation}
\mathbf{V_{tank}}=V_{tank}\mathbf{e_{t}}
\end{equation}

and the force is simply:

\begin{equation}
\mathbf{F_{tank}}=-\mu _{out}\lambda_{t}V_{tank}\mathbf{e_{t}}
\end{equation}

The associated torque
is:

\[
\mathbf{M_{tank}}=\oint_{C}\mathbf{r}\times \mathbf{dF_{tank}}.
\]
The integration provides us with:
\begin{equation}
M_{tank}=-\mu_{out}V_{tank}(L_{1}+L_{2}) 
\label{Mtank}
\end{equation}

where

\[
\begin{array}{c}
L_{1}=\lambda_{t}\oint_{C}(\frac{b}{a}x^{\prime}{}^{2})/\sqrt{\left[(\frac{a}{
b })^{2}x^{\prime}{}^{2}+(\frac{b}{a})^{2} y^{\prime }{}^{2}\right]}dl
\\
L_{2}=\lambda _{t}\oint_{C}(\frac{a}{b}y^{\prime 
}{}^{2})/\sqrt{\left[(\frac{a}{b 
})^{2}x^{\prime}{}^{2}+(\frac{b}{a})^{2} 
y^{\prime}{}^{2}\right]}dl
\label{L1L2}
\end{array}
\]
\end{itemize}
Because the inertial effects are small (and thus neglected), the sum of the 
three torques must vanish. Summing up the three contributions (Eqs.
(\ref{shear}), (\ref{Mtumble}) and (\ref{Mtank})), one finds the evolution
equation for the angular velocity
 of the vesicle: 
\begin{equation}
\frac{d\Psi}{dt}=\omega_{rot}+\omega_{c}+\omega_{elong},
\label{dPsi2}
\end{equation}
where we have defined the three quantities on the r.h.s. of 
(\ref{dPsi2}) as

\begin{equation}
\omega_{rot}=-\frac{\gamma}{2}
\label{vrot}
\end{equation}
 where $\omega_{rot}$ represents the rotational velocity ({\it i.e.} a torque in the Stokes framework), arising from the rotational
part of the flow and is responsible for the global rotation of the 
shape.

\begin{equation}
\omega_{c}=V_{tank} \frac{L_1+L_2}{L_{s1}^2+L_{s2}^2}
\label{vc}
\end{equation}
$\omega_{c}$ is the contribution of the tank-treading motion of the 
membrane to
the effective angular velocity $\frac{d\psi}{dt}$ of the vesicle.

\begin{equation}
\omega_{elong}=-\frac{\gamma}{2} 
\frac{L_{s1}^2-L_{s2}^2}{L_{s1}^2+L_{s2}^2}
\cos (2\psi)
\label{elong}
\end{equation}
$\omega_{elong}$ is the effective elongational velocity which represents
the main contribution of the elongational flow and tends to orient the
vesicle along a direction making an angle $\psi=\frac{\pi}{4}$ with
respect to the applied flow.

It is interesting to note at this point that for a sphere, the various
integrals can easily be computed

\begin{center}
$L_1+L_2 =2\pi \lambda_t a^2$; $L_{s1}^2+L_{s2}^2= 2 \lambda_t \pi
a^3$
\end{center}
Reporting into (\ref{vrot})-(\ref{vc}), and requiring in (\ref{dPsi2}) 
that $d\Psi/dt=0$ (since for a sphere the contribution to tumbling vanishes 
\footnote{Distinguishing between tumbling and tank-treading for a sphere 
might seem a bit confusing. The case of a sphere is degenerate, since one can view 
the dynamics as being of pure tank-treading or pure tumbling nature. Indeed
requiring either that $\frac{d{\Psi}}{dt}=0$, as we did here, or 
$V_{tank}=0$
, provides the same  velocity along the contour.
For continuity reasons with the case where there is a slight deviation 
from
a sphere, we interpret the motion under question as being of 
tank-treading
type.}), we obtain:

\begin{equation}
V_{tank}=\frac{\gamma}{2}a  \label{circle}
\end{equation}
This is the expected value of the tank-treading velocity in the case of 
a
sphere with a radius $a$. Interestingly, this result holds whatever the
prescription for the parameters $\lambda_n$ and $\lambda_t$.

Inspection of eqs (\ref{dPsi2})-(\ref{elong}) reveals, in particular,
that for a rigid membrane where $V_{tank}=0$, no stationnary solution 
is
possible: a rigid elliptic body should always tumble, as one expects. 
This
can be interpreted  by the fact that the rotational velocity $\mid
\omega_{rot}\mid $ is always bigger than the elongational velocity $\mid
\omega_{elong}\mid $. If allowance is made for a tank-treading motion 
(due to the
membrane fluidity and the finite viscosity of the internal liquid) then 
$
\omega_{c}\neq 0$. Equations (\ref{dPsi2}), (\ref{vrot}) and (\ref{vc}) 
show
indeed that the tank-treading motion described by the velocity 
$\omega_{c}$ results
in an effective reduction of the global rotation $\omega_{rot}$, provided 
that the tank-treading
velocity $V_{tank}$ has a positive value. A stationary (non tumbling)
tank-treading motion of the shape is thus possible if the velocity $\mid
\omega_{elong}\mid $, representing the elongational part of the flow, 
can balance the effective tumbling velocity $\mid 
\omega_{rot}+\omega_{c}\mid $. This
can occur for sufficiently high values of the tank-treading velocity.
Hitherto, the tank-treading velocity has been introduced as a 
phenomenological
quantity, and it must be computed independently. This step is necessary 
in
order to have an evolution equation in a closed form. The tank-treading
velocity is clearly limited by the viscous friction of the internal 
fluid,
and this piece of information must be evoked in order to complete the
analysis.

\section{Determination of the tank-treading velocity}

\label{vtank}

Following \cite{keller}, the tank-treading velocity $V_{tank}$ can be
determined by considering the energy dissipated in the system. The 
energy
injected by the flow is dissipated by viscous friction in the fluid 
inside the vesicle. The energy rate (or power) provided by the fluid to an
elementary segment in the laboratory frame is equal to
$\mathbf{dF}.\mathbf{V_{t}}$.

The velocity $\mathbf{V_{t}}$ at a point $\mathbf{r}$ of the membrane
can be
written in the same frame as:

\begin{equation}
\mathbf{V_t}= V_{tank}\mathbf{e_t} + \mathbf{w}\times\mathbf{r}
\label{Vtt}
\end{equation}
where $\mathbf{w}=\frac{d\psi}{dt}\mathbf{ez}$ is the instantaneous 
angular velocity of the vesicle.

An elementary force $\mathbf{dF}$ acting on an element $dl$ of the 
membrane
can be decomposed according to the previous section as:
\[
\mathbf{dF}=\mathbf{dF_{shear}}+\mathbf{dF_{tumble}}+\mathbf{dF_{tank}}
\]
The sum of the torques applied on the vesicle is equal to zero,
entailing:
\begin{equation}
\oint\mathbf{dF}.(\mathbf{w}\times\mathbf{r})dl=\oint\mathbf{w}.(\mathbf{r}\times\mathbf{dF})dl=0
\end{equation}
Hence, only the first contribution of the velocity in (\ref{Vtt}) 
matters.
The total power provided by the flow has the following contributions:

\begin{equation}
E_{tot} = E_{shear}+E_{tumble}+E_{tank}  \label{etot0}
\end{equation}

\begin{itemize}
\item The contribution from the simple shear flow is $E_{shear}=
\oint\mathbf{dF_{shear}}.V_{tank}\mathbf{e_t}$


and upon integration on the contour, we find:
\begin{equation}
E_{shear} = \mu_{out}V_{tank}\gamma\left[\frac{L_1+L_2}{2}+\frac{L_1-L_2}{2}
\cos (2\psi)\right]
\end{equation}

where the lengths $L_{1}$ and $L_{2}$ have been defined previously 
(eqs.\ref{Mtank}).
Following the spirit of the last section, we write 
$E_{shear}=E_{rot}+E_{elong}$ in order to identify
the contributions from the rotational part of the shear flow 
$$E_{rot}=\mu_{out}V_{tank}\gamma\frac{L_1+L_2}{2}$$ and the 
elongational part of the shear flow 
$$E_{elong}=\mu_{out}V_{tank}\gamma\frac{L_1-L_2}{2} \cos (2\psi)$$
This decomposition will be useful in the discussion of the results.

\item The contribution from the tumbling motion is $E_{tumble}=\oint
\mathbf{dF_{tumble}}.V_{tank}\mathbf{e_{t}}$, yielding:




\begin{equation}
E_{tumble}=\mu _{out}V_{tank} \frac{d\psi}{dt}(L_{1}+L_{2})
\label{tum}
\end{equation}


\item The contribution from the tank-treading motion is 
$E_{tank}=\oint
\mathbf{dF_{tank}}.V_{tank}\mathbf{e_{t}}$, and upon integration one 
finds:

\begin{equation}
E_{tank}=-\mu_{out}V_{tank}^{2} P'
\end{equation}
where $P'=\oint_{c}\lambda_{t} dl=\lambda_{t} P$
\end{itemize}

By using the above results, the total power  (\ref{etot0}) takes the 
form:


\begin{equation}
E_{tot} = \mu_{out}(\alpha V_{tank}^2 + \beta V_{tank})
\label{etot}
\end{equation}

where


\begin{equation}
\alpha 
=\left[\frac{(L_{1}+L_{2})^{2}}{L_{s1}^{2}+L_{s2}^{2}}-P'\right]  
\label{alpha}
\end{equation}

\bigskip

\begin{equation}
\beta =\gamma \left[\frac{L_{2}-L_{1}}{2}-\frac{L_{1}+L_{2}}{2}\frac{
L_{s2}^{2}-L_{s1}^{2}}{L_{s2}^{2}+L_{s1}^{2}} \cos (2\Psi)\right]
\label{beta}
\end{equation}

In the particular case of a circular shape, the total power provided by
the external fluid to the internal one can easily be determined:
 indeed, we have $(L_1+L_2)^2 = 4\lambda_t^2 (\pi a^2)^2$; $
L_{s1}^2+L_{s2}^2=2 \lambda_t \pi a^3$; $P' = 2 \lambda_t \pi a$, 
and $
L_1=L_2 $, $L_{s1}=L_{s2}$. This implies that both coefficients
$\alpha$ and
$\beta$ vanish, and so does the total power. This result is conforting
since
inside a sphere the fluid executes a rigid-like rotation (there is no
dissipation) and thus no energy can be transferred. It is only when the
shape
deviates from a circle (or a sphere in 3D) that dissipation is 
permissible. Note that
we arrived at this result before using any information about  
dissipation in the enclosed fluid, and
this points to a consistency of the model.

The energy dissipated by viscous friction in the volume of the vesicle 
is of
the form:

\begin{equation}
\epsilon =\frac{1}{2} \mu \oint_{S} (\frac{\partial V_i}{\partial 
x_j}+\frac{
\partial V_j}{\partial x_i})^2 ds
\end{equation}
In general, we have to determine the velocity field, which satifies the 
Stokes equations
inside the vesicle and  subjected to boundary conditions at the surface of 
the
ellipse. Our aim is not to determine the velocity field exactly, which 
is
not an easy task in general (and an exact result is the exception rather 
than the rule).
Rather we wish to capture the main
ingredients and remain within a heuristic analysis. For that purpose, it 
will be sufficient to make use of an
approximate solution inferred from simple considerations based on the 
result
relative to a spherical shape. In order to anticipate the main 
ingredient, we shall take
the case of a slightly deformed circle as a reference in order to serve 
as a guide for
our reasoning.
Consider $e=\frac{b-a}{a}$ to be small.
The following velocity field fulfills the prescribed conditions ({\it i.e.} to
be a solution of the Stokes equations in the inner domain of the 
vesicle):
\[
\begin{array}{c}
V_{x^{\prime}} = V_{tank}\frac{y^{\prime}}{b}\\
V_{y^{\prime}} = -V_{tank} x^{\prime}\frac{b}{a^2}
\label{speed}
\end{array}
\]
It must be noted that despite the fact that the velocity is not exactly
constant along the contour, the velocity remains colinear to the tangent 
at the membrane
\footnote{Other prescriptions for the flow could have been used.
In particular:
\[
\begin{array}{c}
V_{x^{\prime}} =V_{tank}\frac{y^{\prime}}{b} \\
V_{y^{\prime}} =-V_{tank}\frac{x^{\prime}}{a}
\end{array}
\]
This flow ensures a constant value for the tank-treading velocity along 
the
contour but the velocity is not colinear to the tangential direction of 
the
contour.}
 and this continues to represent a reasonable approximation. Let us
estimate the energy dissipated in the vesicle. This is given by :

\begin{equation}
\epsilon = \mu_{in}\alpha^{\prime} V_{tank}^2 ,
\label{eps}
\end{equation}
where $\alpha^{\prime}$ is a constant depending on $a$ and $b$:
$\alpha^{\prime} = \frac{1}{2} \pi a b (\frac{b}{a^2}-\frac{1}{b})^2$ 
in
the present case.



Using (\ref{etot}) and (\ref{eps}), we arrive at:

\begin{equation}
V_{tank}= -\frac{f_{3}}{f_2-\frac{\mu_{in}}{\mu_{out}}f_{1}}\gamma 
\cos (2\Psi)
\label{evtank}
\end{equation}
where


\bigskip

\bigskip
\begin{equation}
f_{1}=\alpha ^{^{\prime }}=\frac{1}{2}\pi 
ab(\frac{1}{b}-\frac{b}{a^{2}})
\label{f1}
\end{equation}

\begin{equation}
f_{2}=\frac{(L_{1}+L_{2})^{2}}{L_{s1}^{2}+L_{s2}^{2}}-P'  \label{f2}
\end{equation}

\begin{equation}
f_{3}=\frac{L_{2}-L_{1}}{2}-\frac{L_{1}+L_{2}}{2}\frac{
L_{s2}^{2}-L_{s1}^{2}}{L_{s2}^{2}+L_{s1}^{2}} 
\label{f3}
\end{equation}

\bigskip

As could be anticipated, the tank-treading velocity is directly 
proportional
to $\gamma$ which fixes the time-scale of the imposed flow. We also note
that the tank-treading velocity enjoys the same symmetry as the 
elongational
flow does: it vanishes for $\psi=\pm\frac{\pi}{4}$ and is maximal for
$\psi=0$.

Some remarks are in order.
As explained in section \ref{moment}, a shear flow can always be split into
a
rotational part and an elongational one (see Fig. \ref{dshear}). In 
order to understand
the origin of the tank-treading motion, it is appropriate to specify the 
role of both
the elongational and the rotational components of the flow.

On the one hand, for a purely rotational flow, eq. (\ref{etot}) shows 
that
$E_{tot} = \mu_{out}\alpha V_{tank}^2$ since $\beta=0$. Equating
(\ref{etot}) and (\ref{eps}) leads to the condition $V_{tank}=0$,
provided that the shape is not circular. This corresponds to a global
solid-like rotation. On the other hand, for a purely elongational flow, 
eqs.
(\ref{etot}) and (\ref{eps}) lead to $V_{tank} \sim \cos (2\psi)$. A
non-zero tank-treading velocity is possible with the proviso that the
orientation angle is different from $\psi=\frac{\pi}{4}$ 
($\cos(2\psi)\neq 0$). The torque applied on the vesicle arising from the
 elongational flow is (see eq.\ref{Melong}) $M_{elong}\sim \cos (2\psi)$.
An inspection of the balance of
the torques for the elongational flow, as  done in section
\ref{moment}, leads to the dynamical equation $\frac{d\psi}{dt} \sim 
\cos (2\psi)$.
$\psi=\frac{\pi}{4}$ is thus, the only steady equilibrium position, 
with a  tank-treading
velocity  equal to zero. Hence, the existence of a tank-treading
motion of the membrane is only a consequence of the coupling between the
rotational and the elongational part of the  flow. The total effect of
the shear can be
interpreted as follows:  the rotational part tends to push the
orientation
angle of the vesicle axis towards lower values than
$\psi=\frac{\pi}{4}$. As
soon as this is achieved the vesicle acquires a non-zero tank-treading
velocity since there, the elongational part enters into action (see
eq.(\ref
{elong})) \footnote{ Note that the tank-treading velocity is the result
of an energy balance which involves the coupling between the rotational
and the elongational parts of the flow. Since energetic quantities are
quadratic functions of the velocity field, the tank-treading velocity
is not simply the sum of the tank-treading motions associated respectively to
the elongational component and the rotational component considered independently.
Such a summation would result in a vanishing tank-treading velocity}

\section{Dynamical equation for the orientation angle}

Plugging eq. (\ref{evtank}) into eq. (\ref{dPsi2}), we can express
 explicitly $\omega_c$ (which involves the tank-treading motion of the
vesicle) and
this leads to the  general dynamical equation for the orientation angle
$\psi$ of the vesicle:

\begin{equation}
\frac{d\psi}{dt} = A + B \cos (2\psi)
\label{dynamic}
\end{equation}
where

\begin{equation}
A= -\frac{\gamma}{2}
\end{equation}

\begin{equation}
B=-\frac{\gamma}{2} \left[\frac{L_{s2}^2-L_{s1}^2}{L_{s1}^2+L_{s2}^2}
+ (L_2+L_1)\frac{(L_2-L_1)-(L_1+L_2)\frac{(L_{s2}^2-L_{s1}^2)}{(L_{s2}^2+L_{s1}^2}}
{(L_1+L_2)^2-(P'+\frac{\mu_{in}}{\mu_{out}}\alpha^{\prime})(L_{s1}^2+L_{s2}^2)}\right]
\end{equation}
A purely tank-treading motion corresponds to the situation where the
inclination angle is constant. This is expressed by $\frac{d\psi}{dt}=0$
which implies the condition:$-\frac{A}{B}<1$.
This constraint leads, in particular, to a condition on the viscosity 
ratio
between the inner and the outer fluid:

\begin{equation}
\frac{\mu_{in}}{\mu_{out}} < \frac{1}{\alpha^{\prime}}
\left[(L_1+L_2)(1+\frac{
L_{s1}^2}{L_{s2}^2})\frac{L_2}{L_{s2}^2} - P'\right]
\label{tumbling_cond}
\end{equation}
This is the general condition which can be tabulated numerically, 
provided
that the two drag coefficients $\lambda_t$ and $\lambda_n$ are known, 
which
is exactly the case for several shapes (disks, ellipses, spheres). Thus,
 the condition relates uniquely the viscosity contrast to geometrical 
quantities
 which are functions of the swelling ratio.
In order to gain more insight towards an analytical progress we can 
explore the
situation of a small deformation around the spherical shape.
This proves to be sufficient to capture the essential features.
For
that purpose, we set $e=b-a$ and treat $e$ as a small parameter. The 
integrals
$L_1, L_2...$ that enter in (\ref{tumbling_cond}) can be evaluated 
explicitly, so that
the critical condition for tumbling
is expressed in a simple form in terms of the viscosity ratio and the swelling 
ratio $\tau$:














\begin{equation}
\frac{\mu_{in}}{\mu_{out}}=\frac{5\lambda_{t}a}{1-\tau}
\end{equation}

and the dissipation rate (\ref{eps}) scales as:
\begin{equation}
\epsilon=\mu_{in}V_{tank}^{2}(1-\tau)
\end{equation}

This law for the dissipation rate is in a good agreement with previous
numerical results (see \cite{beaucourt03}). We find here that the closer 
is the shape to a sphere, the more difficult does tumbling occur; the 
viscosity contrast for tumbling diverges as $1/(1-\tau)$.

\section{Quantitative and qualitative analyses of the model}

\label{outcome}
We have seen that the model presented here captures the
essential features and sheds light on the various competing effects that 
fix
the tank-treading and tumbling motions. We may ask the question whether
the
model can be made more quantitative. As stated before, the model
requires
the introduction of two drag parameters $\lambda_t$ and $\lambda_n$ which are
the proportionality constants relating  the force and velocity in the
normal
and the tangential directions. 
Let us recall that these two drag coefficients describe the effect of the
hydrodynamic interaction felt by a membrane
element.
This involves the geometry of the shape 
around a given element of membrane and these drag parameters are 
a priori function of the position of the element considered on the contour. If one wishes to go beyond a qualitative
discussion we must determine these two drag parameters, which can be made 
in general only numerically. For a sphere
with a radius $a$ moving in a Stokes flow, it is known that on a local segment of the spherical contour we have: $\lambda_{t}=
\lambda_{n}= 3/2 a$ (see \cite{Brenner}). 
 Instead of evaluating the exact values of these parameters for each elementary elements of the membrane, we shall rather estimate them 
from the best fit with the full
numerical simulations obtained previously(\cite{beaucourt03}).
Note that  each membrane portion can be approximated locally as an arc of a circle. Thus, as in the case of a sphere we chose equal values for the drag coefficients in the normal and in the transverse directions with respect to the contour: $\lambda_{t}=\lambda_{n}$. 
We consider now a vesicle with an aspect ratio $\tau=0.8$ and determine several quantities.

\begin{figure*}[tbp]

\begin{center}
\scalebox{.4}{\rotatebox{-90}{\includegraphics*{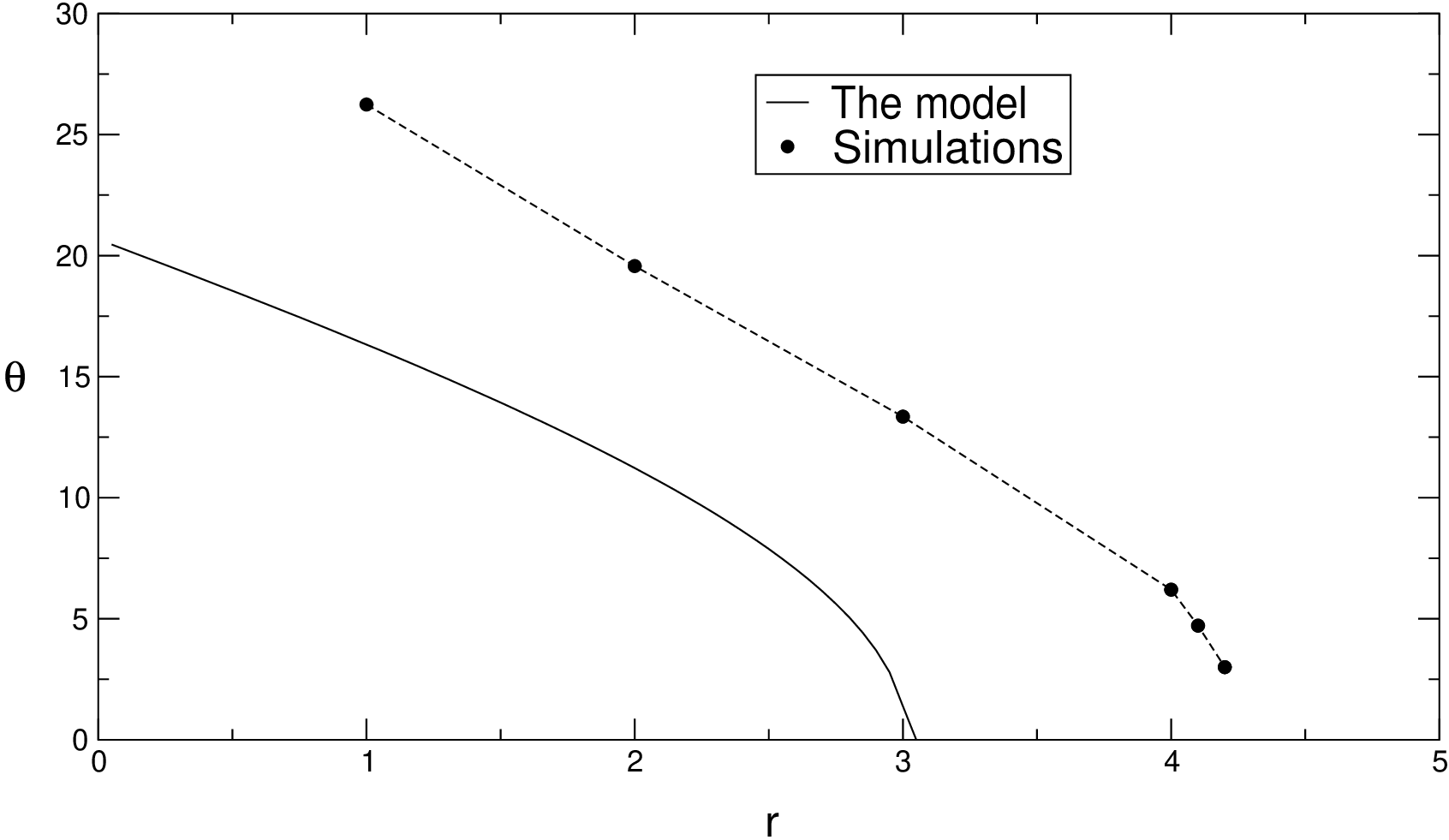}}}
\end{center}
\caption{Equilibrium angle as a function of the viscosity ratio: the
saddle-node bifurcation ($\tau=0.84$: model; $\tau=0.8$:
simulations)}
\label{tetar}

\begin{center}
\scalebox{.5}{\rotatebox{-90}{\includegraphics*{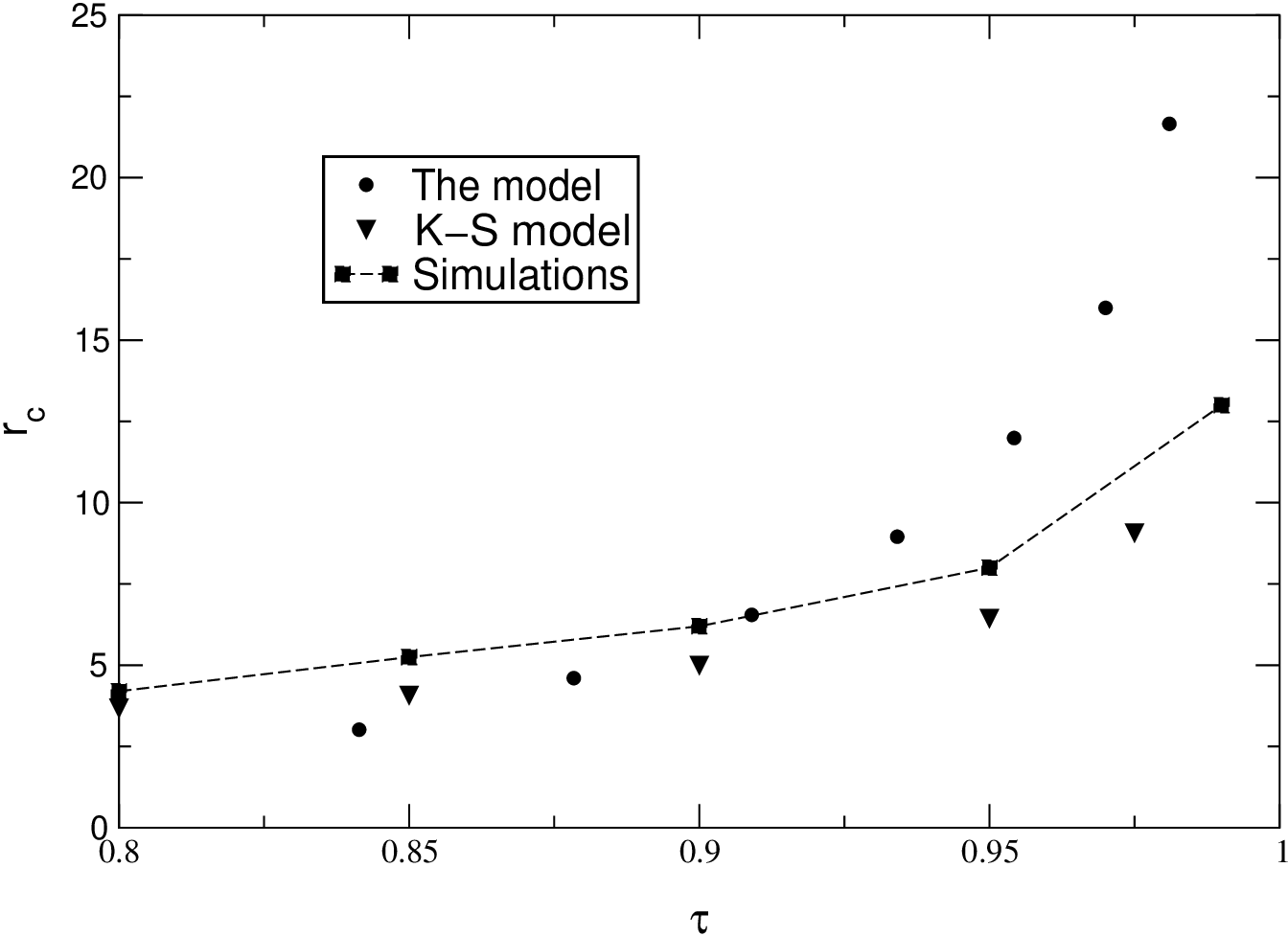}}}
\end{center}
\caption{Evolution of the critical viscosity ratio $rc$ as a function of
 the swelling ratio $\tau$ for $a=1$.}
\label{rctau}
\end{figure*}

\begin{figure*}[tbp]
\begin{center}
\scalebox{0.5}{\rotatebox{-90}{\includegraphics*{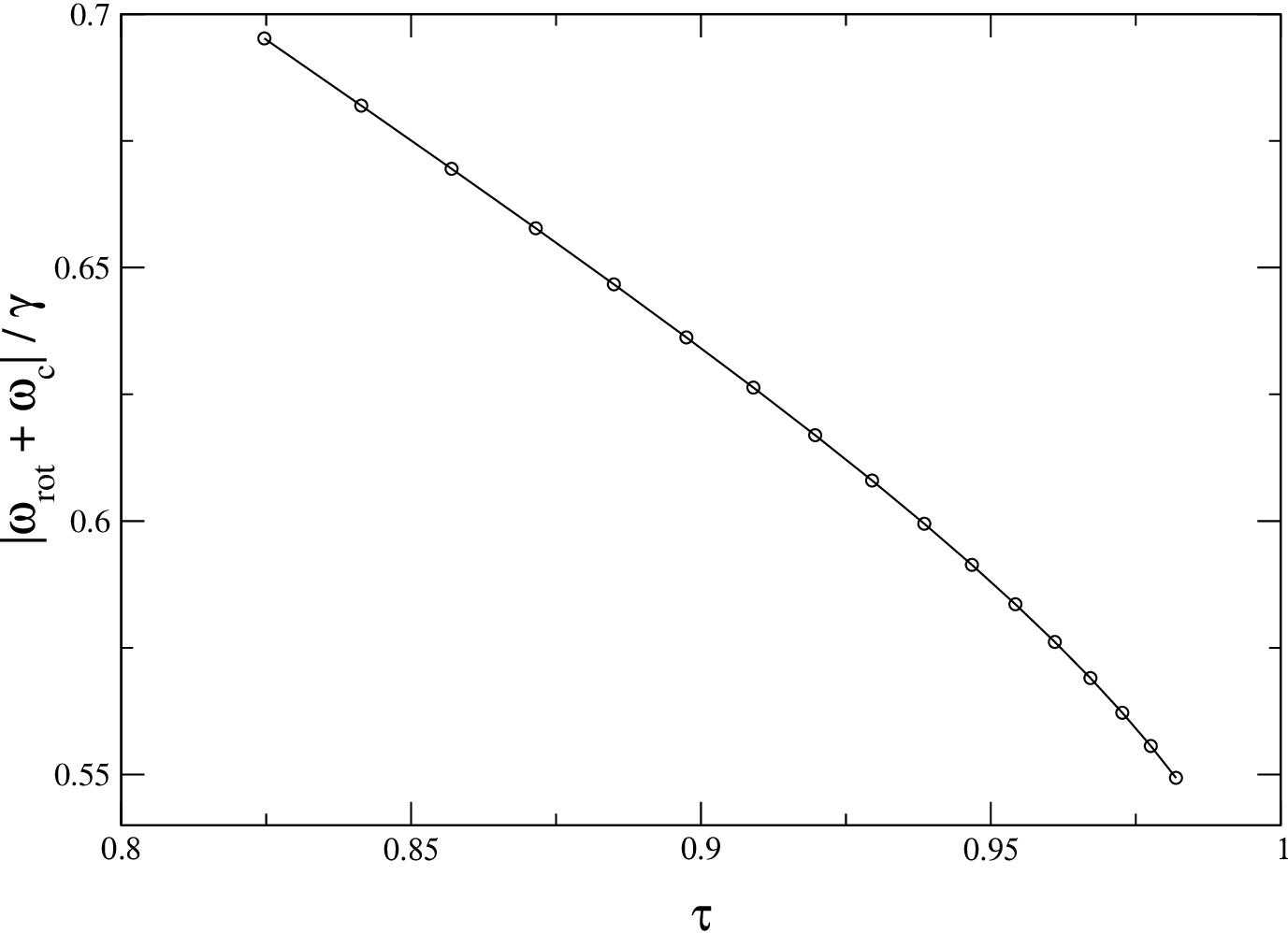}}}
\end{center}
\caption{Evolution of the effective tumbling velocity ($\mid
\omega_{rot}+\omega_{c}\mid/\gamma a$) as a function of the swelling 
ratio $\tau$, $r=\frac{\mu_{in}}{\mu_{out}}=2$}
\label{vrottau}

\begin{center}
\scalebox{0.5}{\rotatebox{-90}{\includegraphics*{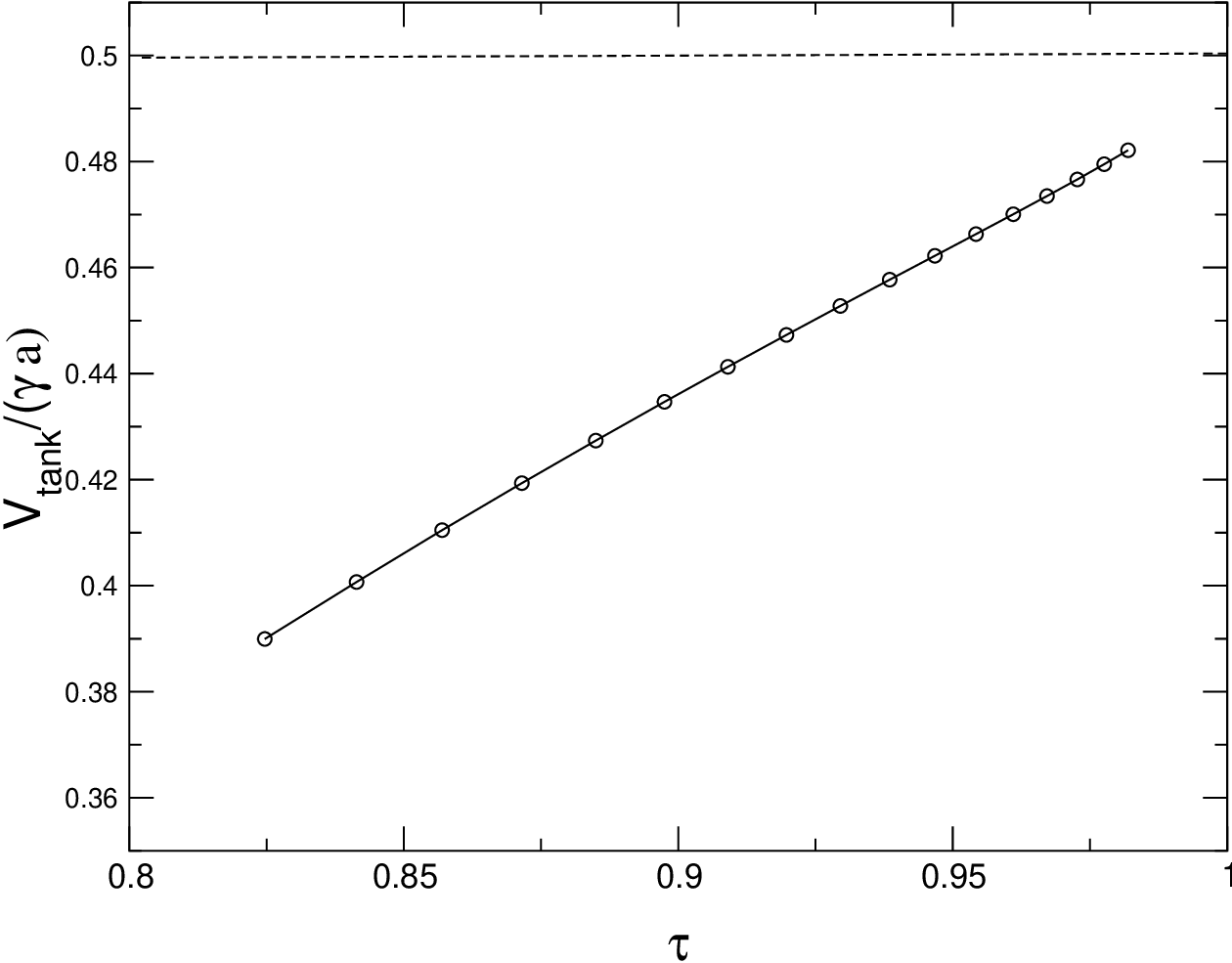}}}
\end{center}
\caption{Evolution of the tank-treading velocity $V_{tank}/\gamma a$ as 
a function of the swelling ratio $\tau$}
\label{vtanktau}
\end{figure*}

\begin{figure*}[tbp]
\begin{center}
\scalebox{0.5}{\rotatebox{-90}{\includegraphics*{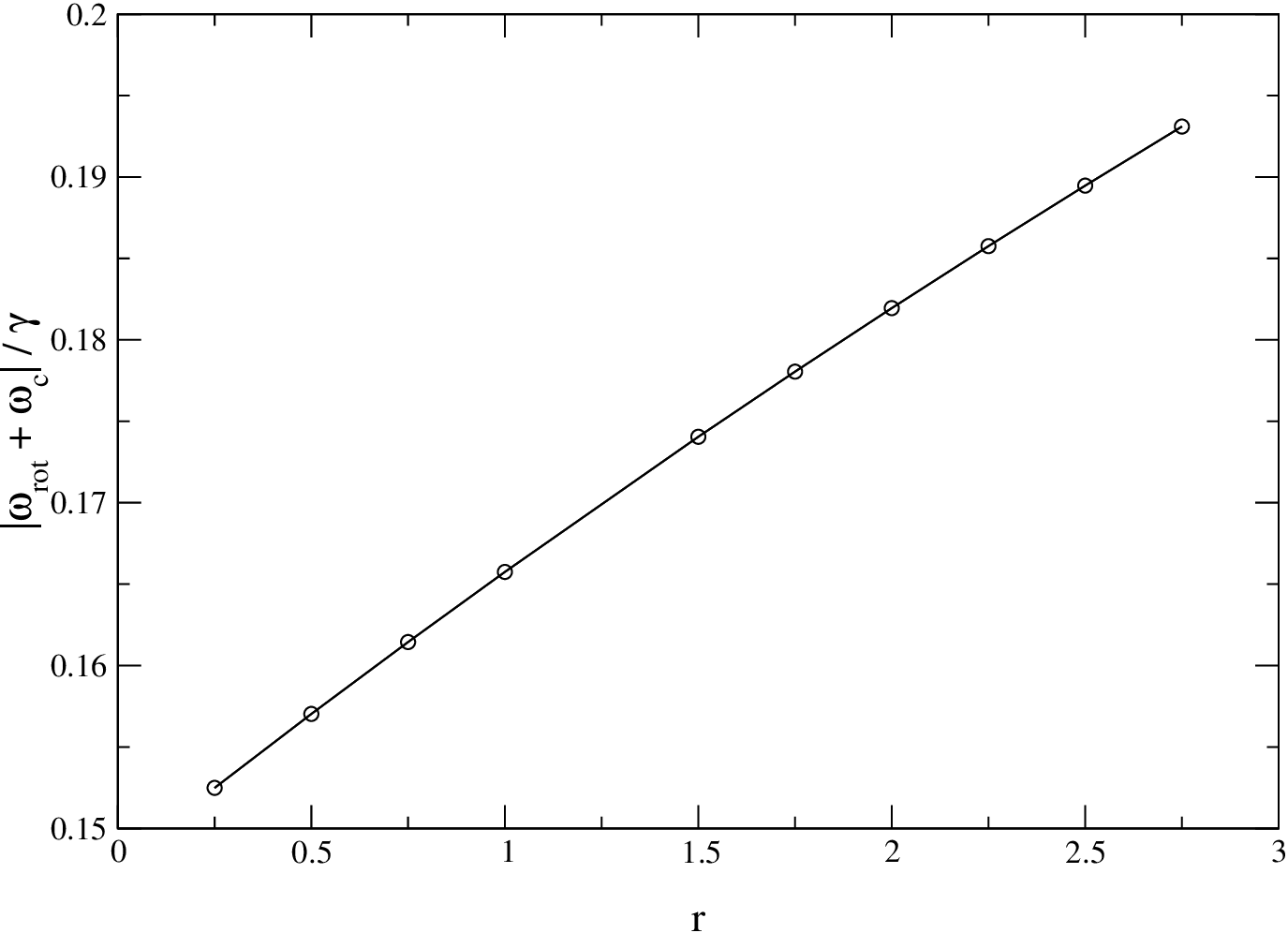}}}
\caption{Evolution of the effective tumbling velocity ($(\mid
\omega_{rot}+\omega_{c} \mid)/\gamma a$) as a function of the viscosity 
ratio $r$, the swelling ratio is equal to $\tau=0.84$}
\label{vrotvisc}
\end{center}
\begin{center}
\scalebox{0.5}{\rotatebox{-90}{\includegraphics*{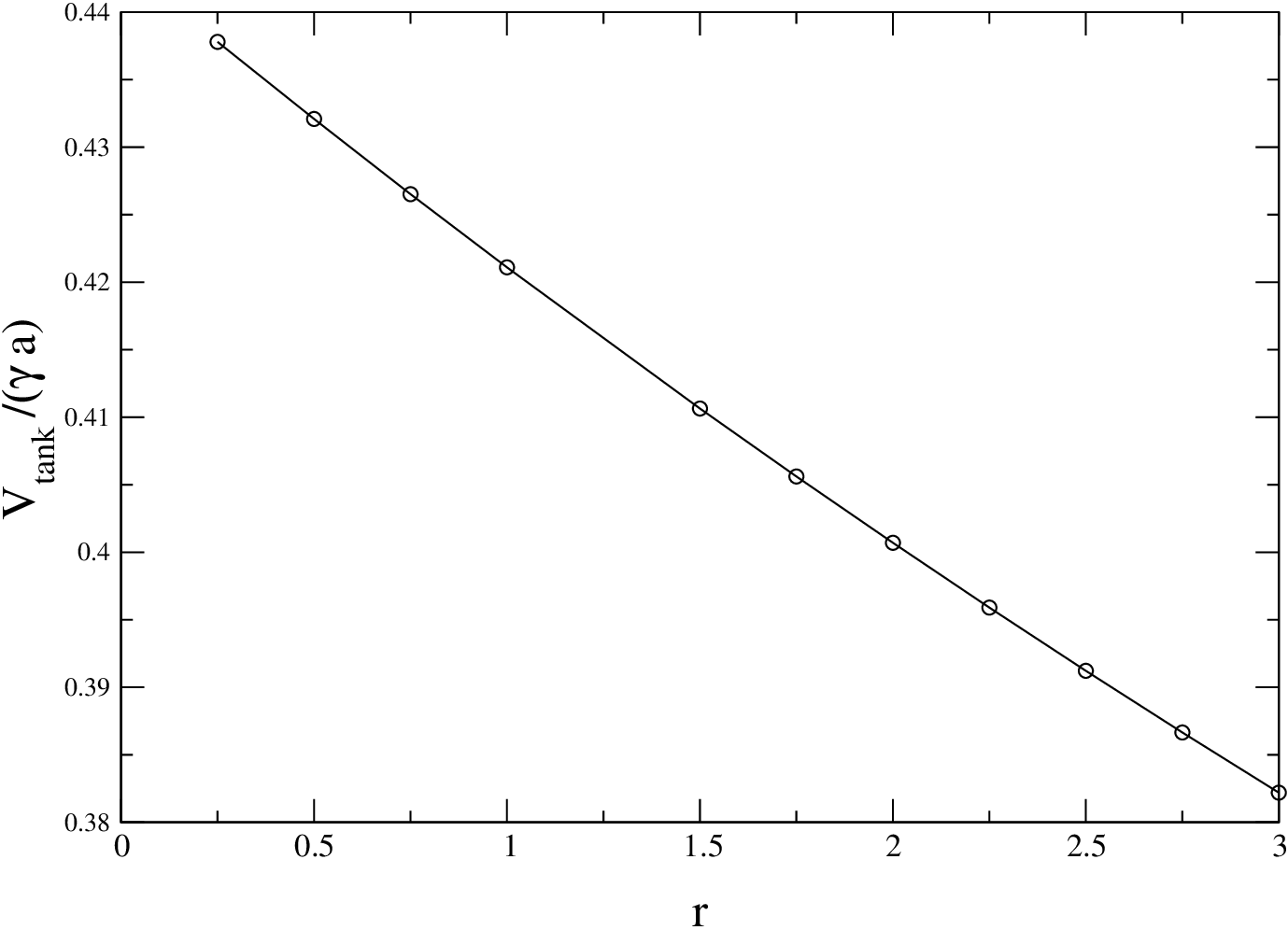}}}
\caption{Evolution of the tank-treading velocity $V_{tank}/\gamma a$ as 
a function of the viscosity ratio $r$, $\tau=0.84$}
\label{vtankvisc}
\end{center}
\end{figure*}

The results have been confronted to the full numerical computation and
the values of $\lambda_t$ and $\lambda_n$ have been guessed. For various
tests made so far, we found
that $\lambda_{t}=\lambda_{n} \simeq 4$ per unit length 
in the transverse direction
provide the most reasonable fit. We present on Fig. \ref{tetar} the evolution 
of the equilibrium angle as a function of the viscosity ratio 
$r$, which corresponds to the stable branch of the saddle-node
bifurcation \cite{biben}. The prediction of the model
qualitatively reproduces the bifurcation branch (this is always the case
regardless of the chosen
parameter), and is fairly in
reasonable agreement on the quantitative level. The point at which the
 angle
is zero corresponds to the threshold of the tumbling bifurcation. This
threshold depends on the swelling ratio. Consequently, the two parameters 
controlling the bifurcation are the viscosity contrast and the swelling ratio. 
Fig. \ref{rctau} represents the boundary between the
 region of
the phase diagram where pure tank-treading motion takes place (low $r$
and high $\tau$) and that,
 where the motion is of tumbling type (which is favored at large $r$ and 
small $\tau$). The results are
compared with the full numerical calculation. It is also worthwhile to
represent some other physical quantities. Of particular interest are the global rotation
velocity and the tank-treading velocity (Figs. \ref{vrottau},\ref{vtanktau} and
\ref
{vrotvisc},\ref{vtankvisc}).

Before concluding some additional comments are worth to mention.
Following
the considerations in section \ref{vtank}, the tank-treading motion
is a result of the competition between the rotational part of the
flow and the elongational component. More precisely, the rotational
component pushes the vesicle axis away from the elongational main axis
($\psi<\pi/4$),  allowing the membrane to acquire a non-zero tank-treading motion.
One may say that part of the rotation torque is transferred to the tank-treading one.
Increasing the viscosity of the inner fluid results in a global reduction of the
tank-treading velocity since the internal dissipation penalizes velocity gradients
inside the vesicle (see Fig. \ref{vtankvisc}). From Fig. \ref{vrotvisc} the effective
tumbling velocity should thus increase, reducing further the value of $\psi$. However,
thanks to the $\cos (2\psi)$ variation of the tank-treading velocity, a new equilibrium
position can be found at a value of $\psi$ which is a decreasing function of the viscosity
ratio. In the extreme limit where $\psi_{eq}=0$, the
elongational
velocity $\omega_{elong}$ reaches its maximum and can not overcome the
rotational velocity
on further increase of the internal viscosity: the steady-state
solution doesn't exist anymore whereby a new dynamical solution takes place in
the
form of tumbling. Fig. \ref{tetar} illustrates the evolution of the
equilibrium angle as a function of the viscosity ratio.

The above discussion was made on the assumption that the swelling ratio is
 constant.
The swelling ratio is a measure of the deviation from a spherical shape.
For the particular case of a circular shape (corresponding to a swelling
ratio $\tau=1$), the total velocity arising from the rotational part
of the
flow is completely transferred in  the tank-treading motion of the 
membrane so that the effective tumbling velocity
$\mid \omega_{rot}+\omega_{c} \mid=0$  (in reality, as commented 
above, this situation is degenerate and there is
no distinction between rigid rotation and tank-treading). For a circular
shape, the tank-treading velocity is maximal and equal to 
$V_{tank}=\frac{\gamma}{2}a$ where $a$ is the radius of the circle. This result has already been obtained directly in section \ref{moment}.

Figures \ref{vrottau} and \ref{vtanktau} represent the evolution of the
tank-treading velocity and the effective tumbling velocity $\mid
\omega_{rot}+\omega_{c} \mid$ as a function of the swelling ratio 
$\tau$. As the shape
deviates from a circular one, the effective velocity responsible for 
tumbling $\mid \omega_{rot}+\omega_{c} \mid$ increases (Fig. \ref{vrottau})
since the tank-treading velocity decreases (Fig. \ref{vtanktau}). This explains 
that the transition to a tumbling regime can be achieved for lower values of 
the viscosity ratio as the swelling ratio decreases. This is indeed what is
observed on Fig. \ref{rctau}.

\section{Conclusion}

We have presented a simple model bypassing the calculation of the Stokes
flow. We have captured the essential features of the transition 
tank-treading/tumbling, and have a
transparent view of the various competing phenomena. This work has added 
a piece to our understanding 
of tumbling.
There are several
important effects that have been disregarded, however.  We have restricted most
of our discussion to 2D shapes. In view of the result of Keller and Skalak\cite{keller} we do not
expect a qualitative change when 3D shapes are considered, provided that the shape
is prescribed. If the shape is free to evolve,  the shear may  induce
a shape transformation, like
a prolate/oblate transition, and this constitutes an important task for future investigations.
In addition, we did not
include the fact that the two monolayers forming the vesicle membrane 
may slide with respect to each other \cite{Seifert93}. In that case one has 
to include two tank-treading velocities, one for each layer, and evaluate
the membrane internal dissipation. It will be an interesting point
to clarify the influence of this fact on tumbling. For 
biological cells, like red cells, further refinement of the model is clearly 
necessary. For example, red blood cells tumble in vivo in the same manner as 
vesicles do. There is however a notable difference between vesicles and red cells. 
The transition to tumbling depends on the shear rate \cite{Goldsmith}. This 
dependence is completely absent for vesicles since there is
 only one time scale $1/\gamma$ which  is imposed by the flow. This 
points to the fact that there should
exist a relevant
intrinsic time scale for red cells. A natural candidate is the elastic
(or even viscoelastic) response of the cytoskeleton.
A natural time scale is
$\mu/G$ where $G$ is the (2D) shear modulus of the spectrin network forming 
the red cell cytoskeleton, and $\mu$ is the  membrane viscosity. Available
data on $G$ \cite{Evans76} and $\mu$ \cite{Dimova} provide us with 
$\mu/G$ of the order of $10^{-2}-10^{-1}$
 seconds, which is not far from $1/\gamma$ in ordinary experiments \cite
{Goldsmith}. We are presently using a simple model for elasticity in 
order
to analyse the qualitative features of this effect \cite{Rioualmisbah}.
\newline
\newline
\textsl{This work has benefitted from a final support from CNES (Centre 
National d'Etudes Spatiales).}

\end{document}